\documentclass[aps,prl,twocolumn,superscriptaddress,a4paper,
longbibliography, amsmath, amssymb]{revtex4-2}

\usepackage{graphicx}
\usepackage{hyperref}
\begin{document}

\title{Practical Quantum Sensing with Thermal Light}

\author{Peng Kian Tan}
\email{cqttpk@nus.edu.sg}
\affiliation{Centre for Quantum Technologies, 3 Science Drive 2, Singapore 117543}
\author{Xi Jie Yeo}
\affiliation{Centre for Quantum Technologies, 3 Science Drive 2, Singapore 117543}
\author{Alvin Zhen Wei Leow}
\affiliation{Centre for Quantum Technologies, 3 Science Drive 2, Singapore 117543}
\author{Lijiong Shen}
\affiliation{Centre for Quantum Technologies, 3 Science Drive 2, Singapore 117543}
\author{Christian Kurtsiefer}
\affiliation{Centre for Quantum Technologies, 3 Science Drive 2, Singapore 117543}
\affiliation{Department of Physics, National University of Singapore, 2 Science Drive 3, Singapore, 117542}

\date{\today}

\begin{abstract}
  Many quantum sensing suggestions rely on temporal correlations found in
  photon pairs generated by parametric down-conversion. In this work, we show
  that the temporal correlations in light with a thermal photon
  statistics can be equally useful for such applications. Using a
  sub-threshold laser diode as an ultrabright source of thermal light, we
  demonstrate optical range finding to a distance of up to 1.8\,km.
\end{abstract}

\maketitle

{\em Introduction --}
Quantum sensing uses quantum phenomena to improve the measurements of
physical parameters  and can be implemented in photonic, atomic or solid-state
systems \cite{pirandola:18}. Photonic quantum sensing techniques include ghost
imaging and  super-resolution imaging \cite{moreau:19}.
 Many photonic quantum sensing
 schemes rely on photon pairs generated in spontaneous parametric down-conversion
 (SPDC)~\cite{ghosh:87} that can be entangled in several degrees of
 freedom, but most often make use of the temporal correlation between the
 photons~\cite{clark:21}. Examples are range finding \cite{lopaeva:13,
   barzanjeh:15} and clock  synchronization \cite{ho:09, jianwei:19} schemes,
 where quantum light sources have an advantage of being stationary, and
 therefore carrying no obvious timing structure that may be subject to
 manipulation or eavesdropping.
 The luminosity of SPDC-based photon pair sources is typically limited to
 sub-nanowatts. 

 In this work, we consider thermal light an alternative resource of
 time-correlated photons by utilizing its photon bunching property. We
 demonstrate quantum sensing using a relatively simple thermal light source
 based on a sub-threshold diode laser.
 The resulting spectral density of this source exceeds that of  SPDC sources by
 approximately 10 orders of magnitude.

{\em Time-correlated photon pairs --}
Photonic sensing applications often make use of modulated light sources and
seek for correlations of a returned signal with the modulation. In an attempt
of moving to low light levels, one can make use of inherent temporal
correlations found in photon pairs emerging from  SPDC in three- or
four-wave mixing processes. These processes generate pairs of photons that
exhibit a strongly peaked second order correlation function
$g^{(2)}(\tau)=f(\tau/\tau_c)$, which characterizes a probability to observe a
pair at a time separation $\tau$. The function $f$ is strongly peaked around
$\tau=0$ (with a spread on the order of a coherence time $\tau_c$), and can be
observed in specialized quantum light sources. Sensing applications based on
this effect are carried out by measuring detection time differences between
one photon acting as a reference, and the other one acting as a probe.

A more natural type of light is thermal light. Thermal light, such as blackbody radiation, exhibits a
characteristic temporal photon bunching behavior \cite{glauber:63a,
  glauber:63b}, also known as the Hanbury-Brown--Twiss effect
\cite{hbt:56a}. This can also be described by a peaked second-order timing correlation,
\begin{equation}
\label{eqn:g2}
	g^{(2)}(\tau) = 1 + e^{-2|\tau|/\tau_c}\,,
\end{equation} 
where $\tau$ is again the timing separation of the two photodetection events,
and $\tau_c$ is the coherence timescale of the temporal photon bunching where
thermal photons have a tendency to be detected closer together than described by
Poissonian statistical timing distribution. Similar to light generated by
SPDC, the coherence timescale $\tau_{c}$ is inversely proportional to the
spectral width $\Delta f\approx c\Delta\lambda/\lambda^2$ of the thermal
light, which is given by the Fourier transform of the source power spectrum
\cite{fox:06}, such that $\Delta f=1/\tau_{c}$ for single-line Gaussian
spectrum. Here, $\lambda$ is the central wavelength of the light,
$\Delta\lambda$ the wavelength spread, and $c$ the speed of light.

\begin{figure}
  \centering\includegraphics[width=\columnwidth]{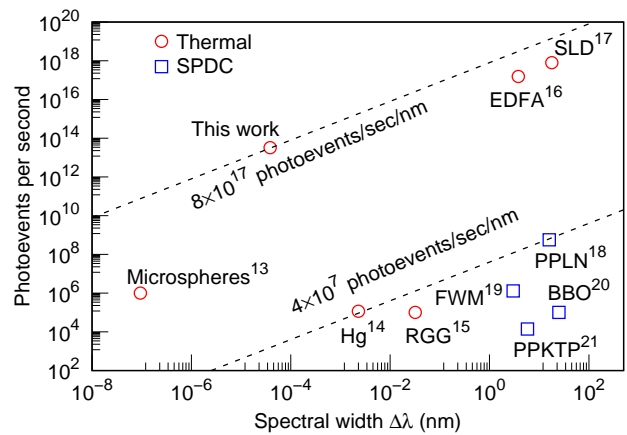}
\caption{\label{fig:comparison} Spectral densities of thermal and SPDC light sources based on: Microspheres\cite{dravins:15} -- suspension of microspheres, Hg\cite{pk:16} -- Mercury discharge lamp, RGG\cite{zhu:12} -- rotating ground glass, EDFA\cite{janassek:18} -- Erbium-doped fiber amplifier, SLD\cite{rahman:20} -- superluminescent diode, PPLN\cite{zhang:15} -- periodically poled Lithium Niobate, FWM\cite{england:19} -- four-wave mixing, BBO\cite{lohrmann:18} -- Beta-Barium Borate, PPKTP\cite{jeong:16} -- periodically poled Potassium Titanyl Phosphate.}
\end{figure}

An important practical consideration for sensing applications is the
brightness of the correlated light source.
As shown in Fig.~\ref{fig:comparison}, SPDC light sources generate an output
power below a nanowatt, or in the range of 10$^{4}$ to 10$^{9}$ photoevents
per second. This
limits the practicality of SPDC light-based sensing in environments with high
attenuation or return loss.
Another important property in a temporal correlation measurement is the
accuracy that can be practically used to infer e.g. a time-of-flight for one
of the photons.
Timing uncertainties of semiconductor-based single-photon detectors are
somewhere below a nanosecond, but more recent nanowire-based detectors may
reach a few picoseconds. When identifying the temporal correlation
feature in thermal light, however, it is necessary that the correlation peak
is still detectable. If the coherence time of the thermal light is
significantly smaller than the detector timing uncertainty, the visibility of
the temporal correlation washes out and may make it impossible to identify it on
top of the Poissonian background. It is therefore desirable to use thermal
light sources with a spectral width below $\approx1$\,GHz.

Thermal light with such a narrow optical bandwidth has been generated in many
different ways. Early examples include single emission lines of gas discharge
lamps~\cite{hbt:56a}. Other methods involve transmitting laser light through random
dispersion media such as suspension of microspheres~\cite{dravins:15}, or a rotating ground
glass plate~\cite{arecchi:65}. These sources, however, have either relatively low output power
due to the spatial incoherence of the randomization mechanism, or (e.g. in the
case of rotating ground glass modulators) a relatively long coherence time.

\begin{figure}
  \centering\includegraphics[width=\columnwidth]{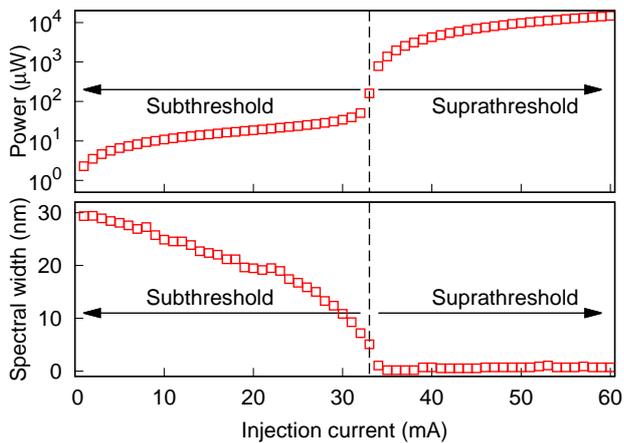}
  \caption{\label{fig:spectra} Output power and spectral width $\Delta \lambda$
    of a laser diode as a function of its injection current to determine the lasing threshold.}
\end{figure}

Here, we use a laser diode operating below the lasing threshold
\cite{bernard:61, lasher:64} to generate thermal light \cite{shockley:52,
  cassidy:91}. This amplified spontaneous emission process generates
significantly higher output power in the range of 10\,$\mu$W to 100\,mW. Light
sources of a similar category include superluminescent diodes, and
Erbium-doped fiber amplifiers. These examples tend to have spectral densities
above milliwatts per nanometer.

The diode laser we use (nominal lasing wavelength $\lambda=518$\,nm, single
spatial mode output) shows a lasing threshold current around 33\,mA
(see Fig.~\ref{fig:spectra}), where the output power exhibits a sharp increase
of 3 orders of magnitude, and the spectrum narrows to a single emission line,
limited by the grating spectrometer to about 0.3\,nm.
We operate the diode laser at 
a subthreshold current of 32.9\,mA, where it
exhibits the photon bunching behavior that is characteristic of thermal
light. The light is then coupled into a single spatial mode optical fiber.
We observe an optical power of $12.5\,\mu$W, corresponding to a photon rate $R\approx3.3\times10^{13}\,$s$^{-1}$, within a spectral window of $\Delta f=43$\,MHz.
This results in a thermal light source of extremely high spectral brightness
of about $8\times10^{17}$ photoevents per second and nanometer.

\begin{figure}
  \centering\includegraphics[width=\columnwidth]{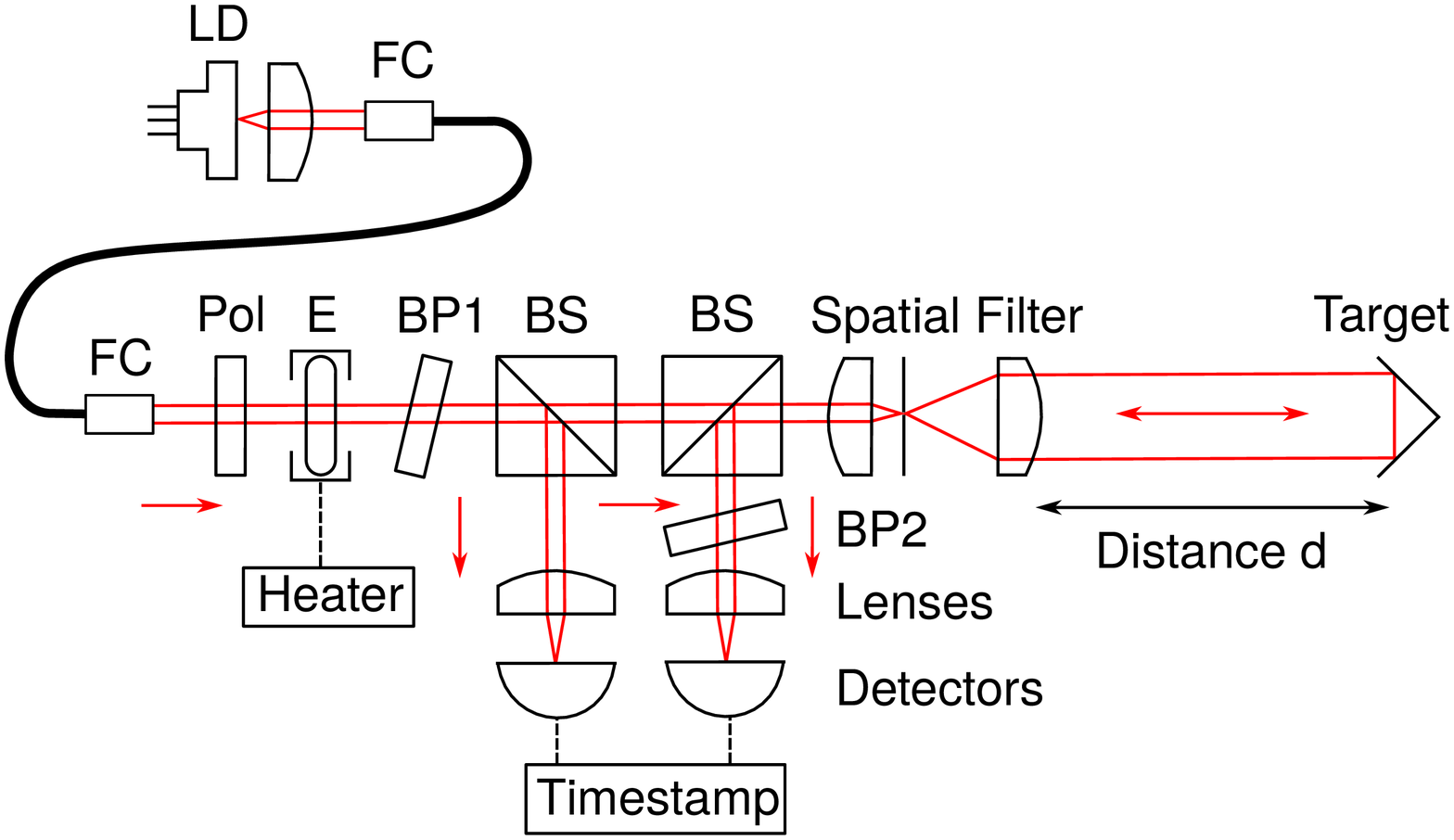}
  \caption{\label{fig:setup}Experimental setup using thermal light for
    ranging measurements. LD: laser diode, FC: Fiber coupler, Pol: polarizer, E:
    etalon, BP1, BP2: bandpass filters, BS: beamsplitter.}
\end{figure} 

{\em Quantum sensing setup --}
The stationary thermal light generated from a subthreshold laser diode is
implemented into an optical ranging setup based on time-of-flight
measurements, commonly known as light distance and ranging (lidar) (see
Fig.~\ref{fig:setup}).
While conventional lidar introduces timing modulation \cite{beheim:86,
  royo:19} into the intensity, amplitude, or phase of the light source, to
provide timing correlations, this work relies on photon bunching of thermal
light to provide the timing correlations. We do record the photodetection
events with a high timing resolution to obtain the temporal photon bunching
signature  $g^{(2)}(\tau)$ similar to \cite{pk:17}.

To ensure that we select only a single chip mode of thermal light from the
sub-threshold laser diode,  a combination of a polarization filter, a bandpass
filter (BP1 in Fig.~\ref{fig:setup}) and a temperature-tuned etalon is used. The
etalon is based  on a fused silica (Suprasil311) substrate and has a tuning
parameter of 4\,GHz/K for its resonances. Optical coatings with a reflectivity
of $97\,\%$ on both sides result in a finesse of 103. The plano-parallel
substrate has a thickness of 0.5\,mm, resulting in a free spectral
range of about 205\,GHz, and spectral transmission windows of 2\,GHz full
width at half maximum (FWHM)~\cite{pk:14}. This allows to effectively to
suppress adjacent laser diode chip modes which are separated by about
50\,GHz from the mode used.
The bandpass interference filter BP1 has a 2\,nm wide
passband centered at $\lambda=518$\,nm to suppress source light beyond the
free spectral range of the etalon.

An asymmetric beamsplitter directs 92\,\% of the filtered thermal light into
the probe beam and retains 4\,\% as a local reference beam sent to a first
single photon detector.

The spectrally filtered thermal light beam passes through a 50:50 beamsplitter
and a telescope formed by a lens pair (f=50\,mm, f=300\,mm)  around a spatial
filter, and is expanded to a diameter of about 50\,mm. The probe beam returns
from the 
target reflector through the same telescope and beamsplitter onto the probe
photodetector. The spatial filter cleans up the returning probe beam and
reduces ambient light contribution from reaching the detectors, and the use of
a second beamsplitter ensures that no breakdown flash light from the target
detector can reach the reference photodetector. A second band pass filter BP2
is used to suppress ambient light reaching the probe detector.

The $g^{(2)}(\tau)$ photon bunching peak is shifted by a time $\tau_0=2d/c$ in
its timing position,
corresponding to the optical path length difference between probe and 
reference beam, thus allowing to infer the distance $d$ of a target
retroreflector from the peak position of $g^{(2)}(\tau)$.

Both single photon detectors are actively quenched Silicon avalanche
photodiodes with a quantum efficiency about $50\,\%$ at 550\,nm, and a timing
jitter around 40\,ps. The detected photoevents are timestamped by an
oscilloscope with (sampling rate 40\,GSPS), and time differences
were histogrammed into 40\,ps wide time bins for short distances $d$, or
recorded with an FPGA-based timestamp device with a resolution of  2\,ns and
sorted into 2\,ns wide time bins numerically for long distances $d$.

\begin{figure}
  \centering\includegraphics[width=\columnwidth]{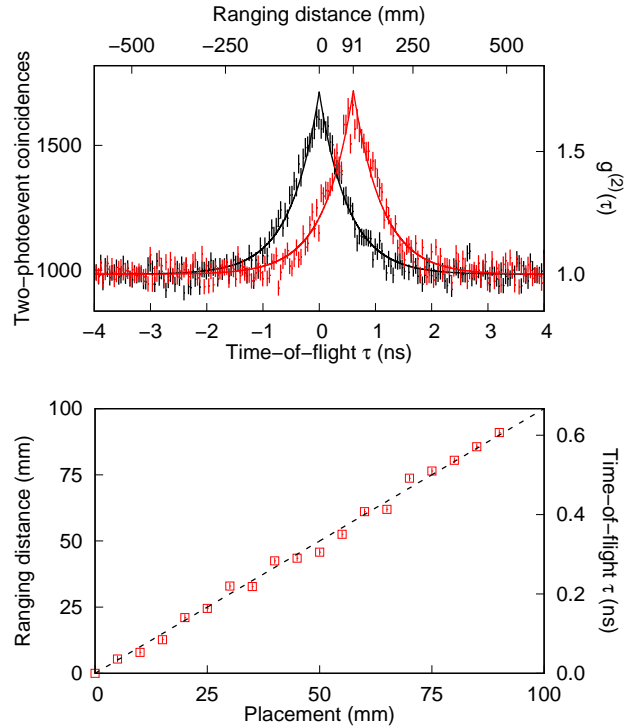}
  \caption{\label{fig:precision} [Top]: Photon bunching $g^{(2)}(\tau)$
    measurements with the target reflector placed at 0\,mm (black) and 90\,mm
    (red). The solid line represents a fit to Eqn.~\ref{eqn:g2} resulting in
    $\tau_{c}=1.03\pm$0.03\,ns, a peak displacement of $\tau_0=0.606\pm$0.008\,ns corresponding to a ranging distance of $d$=91.0$\pm$1.2\,mm, with a
    reduced $\chi^{2}$ of 1.19. [Bottom]: Ranging distances extracted from fits
    to the bunching signatures as a function of the placement positions to test for distance resolution.}
\end{figure}

{\em Range sensing demonstrations --}
Figure~\ref{fig:precision} shows two representative time difference histograms
together with a fitted second-order timing correlation function
$g^{(2)}(\tau)$ according to Eqn.~\ref{eqn:g2}. These allow to determine the
positions $\tau_0$ of their respective bunching peaks and the corresponding
ranging distances $d$ from the round trip time of the probe beam for a set of
target placement positions (Fig.~\ref{fig:precision}, bottom trace). The
resulting ranges are in good
agreement with their corresponding target placement positions, and compatible
with a constraint in the detector timing jitter (about 40\,ps FWHM).

\begin{figure}
  \centering\includegraphics[width=\columnwidth]{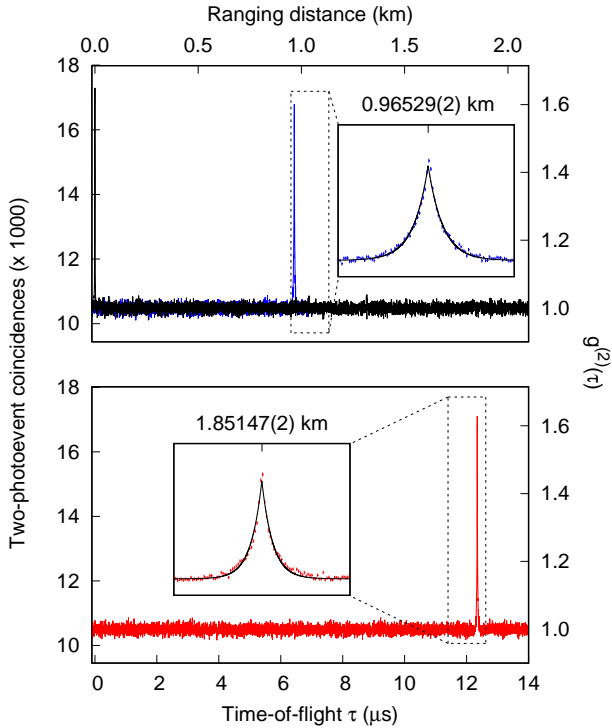}
  \caption{\label{fig:g2} Optical ranging measurements to the reference zero
    distance position with the retroreflector placed at the telescope aperture
    (top, black trace), and the signal obtained with a retroreflector about
    1\,km away (top, blue trace). The bottom trace shows the bunching signature
    with the target retroreflector located about 1.8\,km away from the reference
    detector.}
\end{figure} 

To demonstrate the robustness of the ranging setup, we conducted two outdoor
field measurements. 
Figure~\ref{fig:g2} shows two long-range time-of-flight measurements together
with a reference zero position (black trace), resulting in the
ranging distances of $965.29\pm0.02$\,m (blue) and $1,851.48\pm0.02$\,m (red)
fitted to Eqn.~\ref{eqn:g2} with reduced $\chi^{2}$ of 1.09 and 1.05,
respectively, under the assumption of a unit refractive index of air. The
increased uncertainty compared to the short range measurements shown in
Fig.~\ref{fig:precision} are due to the more coarse histogramming for this
experiment. 

For the long distance  measurements, the etalon temperature tuning and stability was improved relative to the measurements in Fig.~\ref{fig:precision}, increasing the coherence timescale $\tau_{c}=23.2\pm0.4$\,ns (red), corresponding to a spectral linewidth $\Delta f=43$\,MHz. 

The temporal photon bunching peak (red) is slightly reduced to
$g^{(2)}(\tau=0)=1.591\pm0.009$ (red) due to an increase of the bin width from
40\,ps to 2\,ns for the time differences, and by noise contribution from ambient light to the probe detector.

{\em Signal-to-noise considerations --}
The very high spectral density of our light source helps to increase the
signal-to-noise ratio \cite{hbt:74, foellmi:09} of a bunching peak detection
significantly. 
When photodetectors are fast enough to resolve the temporal coherence
$\tau_{c}$ of the photon bunching signature, the signal-to-noise ratio (SNR) of
the second order correlation function  $g^{(2)}(\tau)$ will be dominated by shot
noise of the photodetection events, and can be described by
\begin{equation}
\label{eqn:snr}
	\mathrm{SNR} = r \cdot V^{2} \sqrt{\tau_{c} \cdot \Delta T} \,,
\end{equation} 
with the photoevent rate $r$, the interferometric visibility
$V=\sqrt{g^{(2)}(0)-1}$, the coherence time $\tau_{c}$, and the
integration time $\Delta T$.

With a photon bunching peak value $V^{2}=0.6$ and a coherence timescale
$\tau_{c}=23\,$ns, which corresponds to the measured values in
Fig.~\ref{fig:g2}, an upper bound for the signal-to-noise ratio of around 30
can already be achieved after an integration time $\Delta T=1$\,ms at a
photodetection rate of $r=10^{7}$\,s$^{-1}$ given by typical
avalanche photodetector saturation.
This high tolerance to attenuation
provides the thermal light source an advantage over SPDC light sources in
practical sensing use-cases where significant losses can be expected.

{\em Summary --}
This work explored the use of thermal light for applications where
measurements (like range finding) are based on detecting correlations in time.
Sub-threshold lasers
with their intrinsic temporal correlations thus provide a powerful alternative
to light sources based on spontaneous parametric down-conversion in quantum
sensing applications, and  may offer superior signal-to-noise ratios at a
much reduced system complexity.

With such light sources, a technique originally used for
estimating the size of stars half a century ago can boost a wide range of
practical quantum sensing applications that mostly rely on temporal
correlations.

\begin{acknowledgments}
This research is supported by the Quantum Engineering Programme through
grants QEP-P1 and NRF2021-QEP2-03-P02, the National Research Foundation, Prime
Minister's Office, Singapore.
\end{acknowledgments}


%

\end{document}